%% file: latex/acl_latex.tex
\pdfoutput=1
\documentclass[11pt]{article}

\usepackage[preprint]{acl}
\usepackage{array}
\usepackage{times}
\usepackage{latexsym}
\input{latex/macro}

\usepackage{xcolor}
\usepackage[most]{tcolorbox}
\usepackage{listings}
\usepackage{lipsum}

\lstdefinestyle{graybox}{
    basicstyle=\ttfamily\small,
    breaklines=true,
    breakatwhitespace=true,
    showspaces=false,
    showstringspaces=false,
    numbers=none,
    columns=fullflexible,
    keepspaces=true
}
\definecolor{javagreen}{rgb}{0.25,0.5,0.35} % comments
\definecolor{grey}{rgb}{0.5,0.5,0.5} % medium grey
\usepackage[T1]{fontenc}
\usepackage[utf8]{inputenc}

\usepackage{microtype}

\usepackage{inconsolata}

\usepackage{graphicx}

\title{Butterfly Effects in Toolchains: A Comprehensive Analysis of Failed Parameter Filling in LLM Tool-Agent Systems}

\author{
    \bf{\normalsize Qian Xiong$^{1}$\thanks{These authors contributed equally to this work.}},\ 
    \bf{\normalsize Yuekai Huang$^{2,3}$\footnotemark[1]},
    \bf{\normalsize Ziyou Jiang$^{2,3}$},\\
    \bf{\normalsize Zhiyuan Chang$^{2,3}$},\ 
    \bf{\normalsize Yujia Zheng$^{4}$}, \ 
    \bf{\normalsize Tianhao Li$^{4}$}, \
     \bf{\normalsize Mingyang Li$^{2,3}$\thanks{Corresponding authors.}}\\
  {$^1$ \normalsize Beijing Forestry University} 
  {$^2$ \normalsize Institute of Software, Chinese Academy of Sciences} \\
   {$^3$ \normalsize University of Chinese Academy of Sciences}
   {$^4$ \normalsize Duke University}\\
  \texttt{\normalsize \{x\_qianq\}@bjfu.edu.cn} \\
  \texttt{\normalsize \{mingyang2017,huangyuekai18,ziyou2019,zhiyuan2019\}@iscas.ac.cn} \\ 
  \texttt{\normalsize \{tianhao.li2, yujia.zheng\}@duke.edu} \\ 
 \\
}

\begin{document}
\maketitle
\begin{abstract}
The emergence of the tool agent paradigm has broadened the capability boundaries of the Large Language Model (LLM), enabling it to complete more complex tasks. However, the effectiveness of this paradigm is limited due to the issue of parameter failure during its execution. To explore this phenomenon and propose corresponding suggestions, we first construct a parameter failure taxonomy in this paper. We derive five failure categories from the invocation chain of a mainstream tool agent. Then, we explore the correlation between three different input sources and failure categories by applying 15 input perturbation methods to the input.
Experimental results show that parameter name hallucination failure primarily stems from inherent LLM limitations, while issues with input sources mainly cause other failure patterns. To improve the reliability and effectiveness of tool-agent interactions, we propose corresponding improvement suggestions, including standardizing tool return formats, improving error feedback mechanisms, and ensuring parameter consistency.

\end{abstract}

\section{Introduction}
\begin{figure*}[t]
    \centering
    \includegraphics[width=1\linewidth]{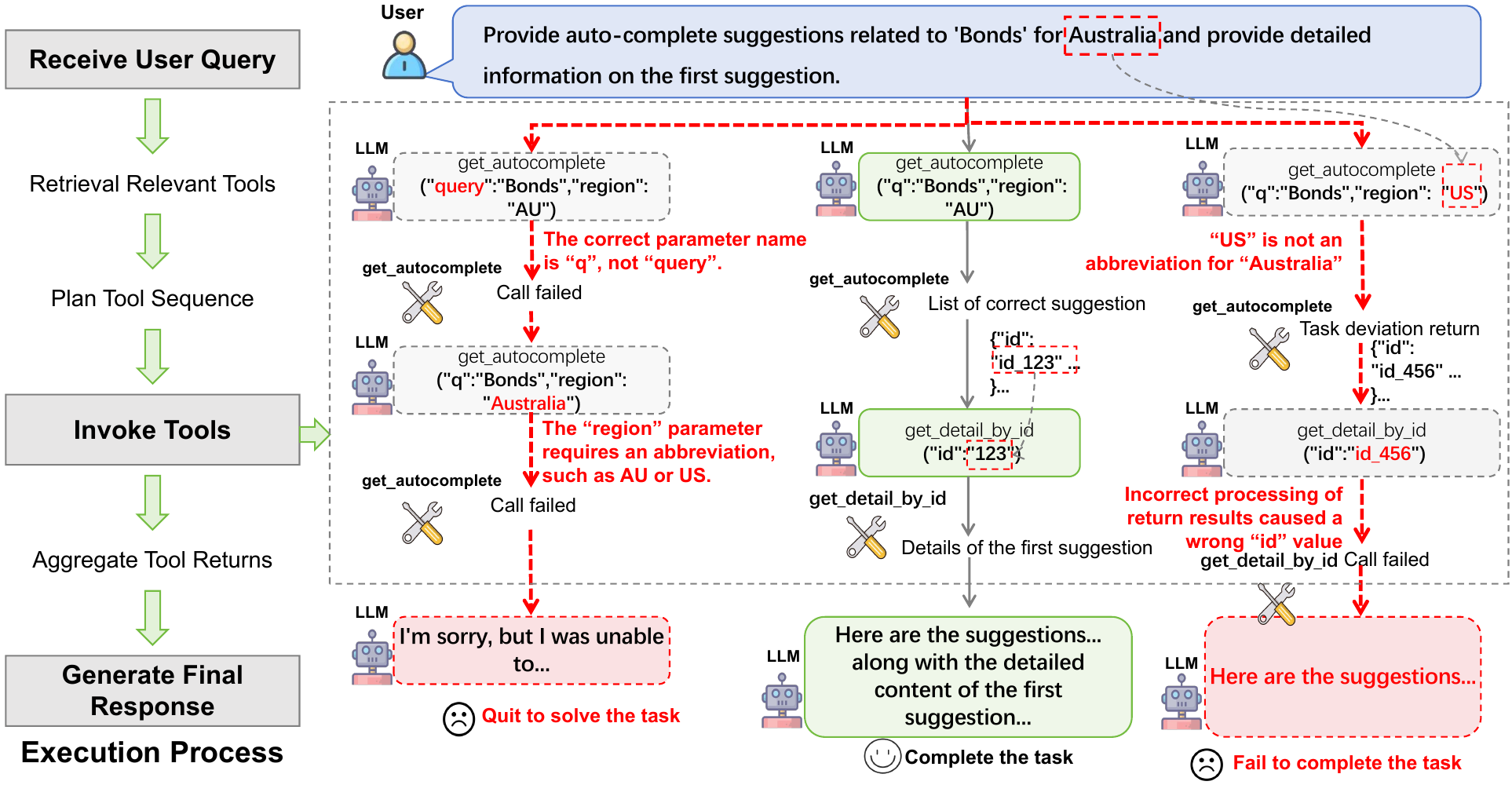}
    \vspace{-2em}
    \caption{The process by which a tool agent resolves a user query, as well as the parameter issues that may occur during the invocation of the toolchain.}
    \vspace{-1em}
    \label{fig:toolagent}
\end{figure*}

In recent years, Large Language Models (LLMs) have shown promising performance in many tasks, but for some professional tasks or tasks that require multi-step processing, relying on a single LLM may not be sufficient to meet the task requirements.
% \rev{The large language models (LLMs) tool agent can leverage external tools to build tool chains through multistep tool invocations, thus completing} 
To address these challenges, researchers have proposed the concept of tool agents, which integrate LLMs with various external tools to complete numerous complex tasks~\citep{Shen2023TaskBenchBL, Qin2023ToolLLMFL, Qu2024ToolLW}. This expands the capabilities of LLMs~\citep{Yao2023LLMLH, Xi2023TheRA} and highlights their significant application potential. 

Figure~\ref{fig:toolagent} illustrates how a tool agent resolves a user query. As we can see, when a tool agent receives a user query, it first retrieves the tool and plans the tool sequence. Then, by parsing the existing input (user query or output result of the previous tool), it forms a parameter list for the subsequent tool invocation. Consequently, the proper execution of tool agents is highly dependent on the accuracy of parameter parsing, which is a critical process. However, parameter errors or parameter hallucinations often occur during the parameter parsing process due to incomplete or ambiguous \textbf{user queries}, low-quality \textbf{tool document} parameter specifications, non-standard \textbf{tool return results}, and limitations of LLM capabilities~\citep{Zhang2024AgentSB, Ye2024ToolSwordUS}. For example, in the erroneous invocation chain on the left side of Figure~\ref{fig:toolagent}, the tool does not call success because ``\texttt{Australia}'' does not meet the abbreviation requirement for the ``\texttt{region}'' parameter. It also includes deviations from the intent of the user. On the right side of the figure, setting the region to ``\texttt{US}'' does not match the country ``\texttt{Australia}'' in the query. More failure cases can be seen in the Appendix~\ref{app: cases}.

As demonstrated in the example, parameter issues often lead to failed tool invocations by the agent and may even confuse the invocation chain, thereby reducing the quality of task completion, like the ``\textit{Butterfly Effect}'' in the toolchains that affects the normal tool agent execution. Moreover, according to existing research~\citep{Zhang2024CrabsCR}, parameter issues commonly exist in the execution traces of tool agents. Data shows that about 44\% of simple user queries and 48\% of complex user queries have parameter issues, which greatly limits the usability of tool agents, especially in some critical domains ~\citep{Cemri2025WhyDM, Ruan2023IdentifyingTR}.
These concerning figures highlight the urgency of addressing the parameter issues in tool agent execution traces.

Many research works have been proposed to explore the above issues~\citep{Singh2024EvaluatingTA, Qin2023ToolLLMFL, Wang2024GTAAB, Li2023APIBankAC}. However, they either focus on specific tasks or emphasize the completion of overall tasks while neglecting the internal logic of tool effectiveness and the mechanisms of parameter handling. In addition, few studies comprehensively consider the impact of different input sources on the accuracy of parameter filling, including user queries, tool documents, and tool return results.

To bridge this gap, we propose an empirical study that analyzes the parameter filling process in tool agents. We aim to construct a comprehensive parameter failure taxonomy, deeply analyze the impact of different input sources on parameter filling accuracy, and provide actionable insights to enhance the reliability of tool agents, alleviating the parameter generation errors and hallucination issues. Specifically, we used the current mainstream tool agent ToolLLaMa~\citep{Qin2023ToolLLMFL} as the investigated target and scientifically applied Grounded Theory (GT)~\citep{Glaser1967} to develop a failure taxonomy through open coding and constant comparative analysis. By applying 15 perturbation methods to input sources, we analyzed the impact on agent parameter behavior on four advanced LLMs.

The contributions of this paper are as follows.
\begin{itemize}
    \item We systematically identify parameter failure patterns and construct a failure taxonomy for API-type tool invocations. 
    \item We propose 15 input perturbation methods and conducted perturbation experiments on three types of input sources involved in the tool agent, revealing the impact of different input sources on parameter failure issues.
    \item We provide practical advice to enhance the effectiveness and reliability of tool agents.
    \item We release the code and dataset\footnote{\href{https://github.com/xQianQ/toolagent-parameter-failure}{https://github.com/xQianQ/toolagent-parameter-failure}} to facilitate further research. 
\end{itemize}

\section{Related Work}
\textbf{Evaluation of LLMs Tool Use}
Evaluating LLMs' tool use abilities has drawn widespread attention.~\citep{Tang2023ToolAlpacaGT, Patil2023GorillaLL, Qin2023ToolLLMFL} focus on the model's tool usage, and~\citep{Huang2023MetaToolBF} assess the understanding of tool usage and tool selection .~\citep {Huang2024PlanningCU} covers a wide range of dimensions and explicitly evaluates the model planning part.~\citep{Li2023APIBankAC} demonstrates the tool-calling capabilities of different models, including planning and retrieval .~\citep {Wang2024GTAAB} focuses on real-world multi-model context inputs and~\citep{Ye2024ToolSwordUS, Ruan2023IdentifyingTR} emphasize the necessity of improving the robustness and security of LLMs in tool use.

The execution process within the tool agent is shown in Figure~\ref{fig:toolagent}. And it typically involves receiving user query tasks, retrieving relevant tools, planning the execution sequence of tools, and invoking the selected tools. Finally, it parses and aggregates the results returned by the tools and generates a final response to the user. In this process, most current research focuses on how to better achieve tool retrieval and selection, paying attention to the final response~\citep{Ning2024WTUEVALAW, Wang2024MTUBenchAM}. However, it fails to delve deeply into the risk points when invoking the tools. Moreover, since existing research does not discuss and analyze the source of parameter information~\citep{Lu2024ToolSandboxAS}, the evaluation work is difficult to reflect the complex input scenarios in the real world. We have discovered that the quality of parameter filling significantly influences whether tools can effectively enhance LLM's problem-solving capabilities. Therefore, our research focuses on the issue that LLMs provide incorrect parameters during the tool invocation phase. Specifically, we performed a comprehensive analysis of LLM parameter-filling behaviors and constructed a data-driven taxonomy. In addition, to further simulate complex real-world scenarios, we strive to perturb the input source of parameter information. By analyzing the changes in the LLM's parameter behavior, we can provide effective suggestions for improving the reliability and effectiveness of the tool agent. 

\input{latex/section/taxonomy}

\input{latex/section/methodology}

\input{latex/section/experiment}

\input{latex/section/results}

\vspace{-1em}
\section{Conclusion}
The study delves into the complex parameter challenges faced by LLM tool agents during the execution of the toolchain, revealing five distinct failure patterns that disrupt workflow integrity. Missing required parameters will prevent the tool from fully processing the task; redundant information affects the accuracy of the tool's return results; hallucination names and mismatch specifications fundamentally prevent the tool from being invoked, and task deviation affects the practical value of the result to the user. Ambiguous user queries, defective tool documents, or poorly written tool return results can spread throughout the toolchain, causing parameter failures and ultimately leading to task failure. This paper provides a blueprint for designing robust tool agents by mapping failure patterns and their interconnections, highlighting proactive input structuring, adaptive error correction, and attention to parameter propagation in the toolchain to ensure reliable and traceable toolchain operations.

\section*{Limitations}
Our experiments focused mainly on data in English. There are significant differences in morphology, syntax, and semantics in different languages. This could influence how parameters are extracted and processed from the input information. Therefore, there may be differences in the failure taxonomy and the relationships between the observed failure patterns. Future research should broaden its scope to include multilingual data, improving the generalizability of our findings. In addition, our research was limited to relatively controlled single-turn conversation scenarios. It lacked consideration of real-time and multi-turn conversation scenarios and mainly revolved around API-type tool invocations. Errors in non-API-type tools, such as command-line tools and libraries used in specific programming language environments, were not explored in this article. Future research should investigate parameter filling processes and failures across a broader range of tool types to develop more comprehensive strategies to improve the reliability of tool agents.

\section*{Ethical Considerations}
The main ethical concern of this article is that the use of the tools may involve personal sensitive information, posing a risk of privacy leakage. Additionally, for some behaviors related to the local operating system environment, a virtual environment isolation mechanism should be set up. Ensure that tool interactions and executions occur within a sandbox environment to prevent the direct exposure of sensitive system details.

\bibliography{acl_latex}

\appendix
\input{latex/section/appendix}
\end{document}

%% file: latex/macro.tex
\usepackage{booktabs}
\usepackage{multirow}
\usepackage{graphicx}
\usepackage[normalem]{ulem}
\usepackage[preprint]{acl}
\usepackage{amsmath}
\usepackage{times}
\usepackage{latexsym}
\usepackage{graphicx}
\usepackage[T1]{fontenc}
\usepackage{amssymb}
\usepackage{booktabs}
\usepackage[utf8]{inputenc}

\usepackage{microtype}

\usepackage{inconsolata}
\usepackage{changepage}
\usepackage{multirow}
\usepackage{array}
\usepackage{makecell}

\usepackage{longtable}

\usepackage{soul}
\usepackage{multirow}
\usepackage{epstopdf}
\usepackage{hyperref}
\usepackage{listings}
\usepackage{fancybox}
\usepackage{graphicx}
\usepackage{color}
\usepackage{paralist}
\usepackage{ragged2e}
\usepackage{enumitem}
\usepackage{bigstrut}
\usepackage[utf8]{inputenc}
\usepackage{booktabs}
\usepackage{bm}
\usepackage{subfigure}

\definecolor{javared}{rgb}{0.6,0,0} % for strings
\definecolor{javagreen}{rgb}{0.25,0.5,0.35} % comments
\definecolor{javapurple}{rgb}{0.5,0,0.35} % keywords
\definecolor{javadocblue}{rgb}{0.25,0.35,0.75} % javadoc

%% file: latex/section/taxonomy.tex
\section{Failure Taxonomy Construction}
\begin{figure*}[t]
    \centering
    \includegraphics[width=1\linewidth]{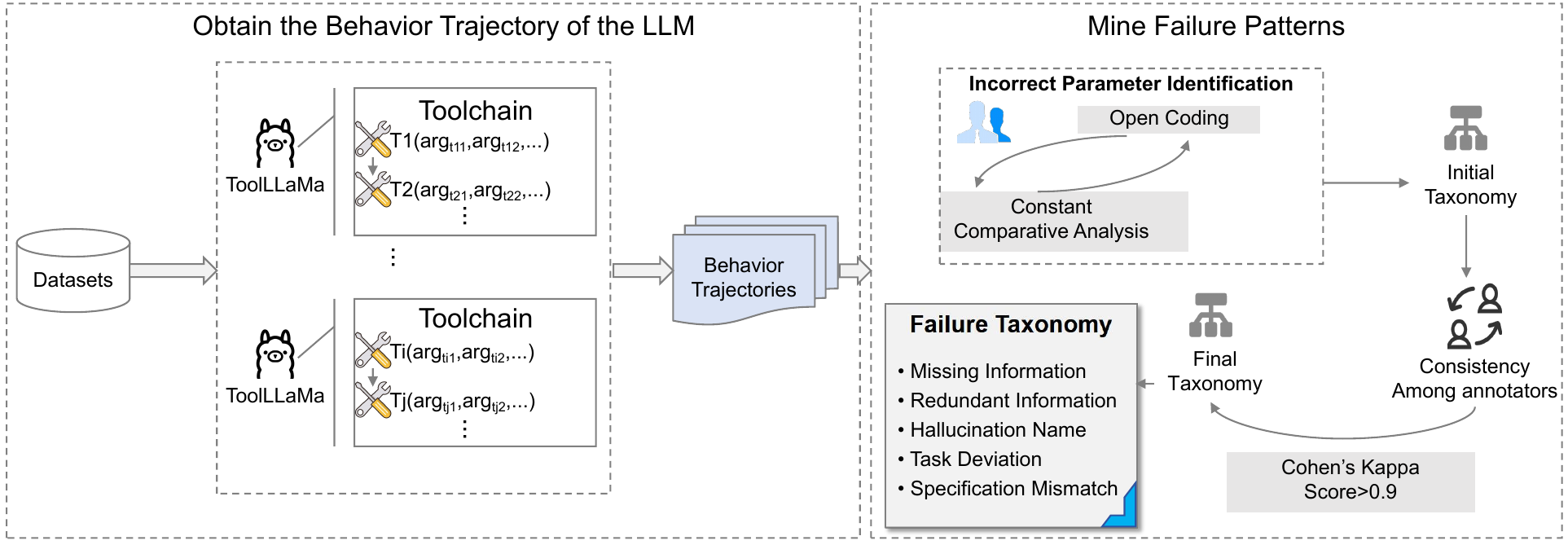}
    \vspace{-2em}
    \caption{The process of systematically mining the failure patterns using the Grounded Theory Approach (GT)~\citep{Glaser1967} to obtain a scientific failure taxonomy.}
    \label{fig: taxonomy}
\end{figure*}
The purpose of this study is to explore the impact of parameter failure issues on tool agents, identify the root causes of tool invocation failure involving parameters, and propose improvement suggestions. However, there is not yet a comprehensive parameter failure taxonomy in place. Therefore, before conducting experiments, we first need to construct a parameter failure taxonomy. The construction process is shown in Figure~\ref{fig: taxonomy}.

We first utilized benchmark datasets reported in ~\citep{Ye2024ToolEyesFE}, and adopted ToolLLaMa as the foundational LLM to obtain the behavior trajectories of the LLM when resolving user queries. The selected dataset and LLM are widely used in the field of tool agent research, ensuring the representativeness of the results.
After obtaining the behavior trajectories, they will be utilized for mining failure patterns. We followed the Grounded Theory Approach (GT)~\citep{Glaser1967}, which is a qualitative research method that directly constructs theories from empirical data, to identify failure patterns.

Specifically, the annotators first understand the function of each available tool and clarify the relevant parameter requirements, then apply the Open Coding approach~\cite{Khandkar2009} to analyze the agent parameter filling behavior in the environment interaction trajectories that we collected. Open coding is a data analysis method that involves decomposing and conceptualizing data to generate corresponding codes for identifying key patterns, themes, or phenomena in the data. Then, using constant comparative analysis, we systematically compared the new codes created by the annotators with the existing codes. This iterative process of parameter failure pattern identification and open coding continued until no new insights emerged from additional data. Eventually, we obtained preliminary patterns from ToolLLaMa's behavioral trajectories.

To refine the patterns, we conducted a group agreement study to ensure that all three annotators had a consistent understanding of the results and gradually made iterative modifications to form the consensus patterns. This process continues until the Cohen’s Kappa score reaches above 0.9, which, as suggested by \citep{8d20e0b8-89d8-3d65-bcf5-8c19d56ec4ab}, indicates an almost perfect level of agreement among annotators and high data quality.
Finally, we constructed a taxonomy that includes five distinct categories of parameter failures that occur during tool agent invocation as follows:
\begin{itemize}
\item \textbf{Missing Information:} This means that when invoking tools, the LLM fails to fill in all the parameters required to solve the task, resulting in the tool obtaining less information than necessary. This not only leads to imprecise results returned by the tool, but also directly causes the failed tool invocation when required parameter items are missing. 

\item \textbf{Redundant Information:} This means that when invoking tools, the LLM sets some additional parameters within the range identifiable by the tool that were not mentioned by the user. For example, it may limit the number of results that the tool returns. Although this case usually does not cause the failed tool invocation, it restricts the range of returned results, thereby affecting the final result.

\item \textbf{Hallucination Name:} This means that when invoking tools, the LLM has raised a parameter name hallucination error, generating parameter items that are not within the tool's recognition range, which prevents the tool from being correctly invoked. This stops the tool from responding normally. In Figure~\ref{fig:toolagent}, the ``\texttt{query}'' parameter set by the LLM for \texttt{get\_autocomplete} is exactly an example of a parameter name hallucination failure. 

\item \textbf{Task Deviation:} This means that the parameter values deviate from the requirements of the target task. For example, inconsistencies in crucial task-related information, such as regions, times, and ID attributes, fall into this category of failure. In such cases, the tool appears to be invoked correctly from a macroscopic perspective, but it actually misleads the LLM, causing it to generate incorrect final solutions to the user's query. 

\item \textbf{Specification Mismatch:} This means that the parameter values set do not match the specifications defined in the tool document. Mismatch in parameter types, value ranges, and formats will prevent the tool from processing the parameters as expected, thereby affecting the normal use of the tool and the quality of task completion.
\end{itemize}

%% file: latex/section/methodology.tex
\section{Methodology}
Based on the constructed taxonomy, we can conduct a corresponding sensitivity analysis on the phenomenon of parameter failure. This process can draw on the ideas of defect detection in the field of software engineering, which leverage test oracle to evaluate the results of the test execution and determine if they meet expectations~\citep{Young2001TestO}
Based on this concept, we treat the initial correct behavior trajectories as test oracles and perturb them to generate the test samples. Considering that there are three input sources (i.e., \textbf{Tool Document}, \textbf{User Query}, \textbf{Tool Return}) of tool agent that will have impact on the parameters, to conduct comprehensive analysis, we design targeted perturbation methods for each of the three input sources to generate corresponding test samples, which will be explained in detail as follows.

\subsection{Tool Document Perturbation}

For the tool document, there are 6 perturbation methods, namely RD (Removed Description), RE (Removed Example), WD (Wrong Description), WT (Wrong Type), SD (Swapped Description), and CO (Changed Order). These methods perturb the parameter description documents from different aspects, and for comprehensive algorithmic descriptions and implementation details of each perturbation method, please refer to Appendix~\ref{app: d methods detail}.
\begin{itemize}
    \item \textbf{RD and RE test the behavior of LLM under information-lack conditions.} The RD removes the description information of the required parameters, while the RE erases all the usage example information. They are intended to investigate whether the model can accurately comprehend and apply parameters in the absence of crucial descriptions or example guidance.
    \item \textbf{WD and WT test LLM sensitivity to environmental noise.} WD substitutes the parameter descriptions in the document with those of other irrelevant tools, and the WT algorithm alters the data types of the parameters. This enables us to understand the influence of such incorrect information on the model's understanding of parameters and its correct utilization of tools.
    \item  \textbf{SD and CO test LLM sensitivity to changes through information order and correspondence.} SD swaps the usage description information of a specified pair of parameters, and CO rearranges the order of the parameter usage descriptions. This helps to understand the model's adaptability when faced with changes in the order of parameter descriptions or the correspondence between descriptions and parameters.
\end{itemize}

\subsection{User Query Perturbation}
For user query, there are 4 perturbation methods, namely \(RP_F\) (Remove First Parameter),  \(RP_L\) (Remove Last Parameter),  \(CP\) (Complicate Parameter), and \(AN\) (Add Noise). These algorithms perturb the original user query \(Q\) from different dimensions, aiming to explore the performance of the model when faced with changes in this input source, and for comprehensive algorithmic descriptions and implementation details of each perturbation method, please refer to Appendix~\ref{app: q methods detail}.

\begin{itemize}
    \item \textbf{ \(RP_F\) and \(RP_L\) test the LLM's dependence on the integrity of user query.} \(RP_F\) removes the first parameter information from the user query, while \(RP_L\) removes the last parameter information. They are used to evaluate whether the model can still execute tasks based on the remaining parameters when key parameters are missing in the user query.
    \item \textbf{ \(CP\) and \(AN\) test the anti-interference ability of the LLM when faced with complex or noisy queries.}  \(CP\) replace the parameters with complex descriptive phrases, while \(AN\) adds interfering information similar to the parameters after the query. This is helpful for testing the adaptability and robustness of LLM in real and variable user query environments. 
\end{itemize}

\subsection{Tool Return Perturbation}
For tool return, there are 5 perturbation methods, namely FK (Fuzz Key), AP (Apply Prefix), CK (Camel Case Key), UK (Underscore Notation Key) and CF (Corrupt JSON format). The purpose is to explore the specific impacts of different types of changes in tool return on the performance of LLM, and for comprehensive algorithmic descriptions and implementation details of each perturbation method, please refer to Appendix~\ref{app: t methods detail}.

\begin{itemize}
    \item \textbf{FK and AP test LLM's understanding and utilization of keys and values in the JSON format tool return.} The FK algorithm replaces the key names in the tool return with ``Object\_i'', where i varies according to the number of keys, eliminating the original semantic information of the key names. Simulates the situation where the semantic information of key names may be ambiguous in practical applications. AP, specifically targeting ID-type return (the most likely to be utilized as parameters in tool chain scenarios), adds the prefix ``ID\_'' to them. By perturbing the semantic information of key names and adding prefixes to ID types of return, we can observe the sensitivity and processing ability when changing the tool's return results.
    \item \textbf{CK and UK test the LLM's compatibility and adaptability to these different naming conventions.} CK converts the key names in the tool return into camel-case notation, while the UK converts them into underscore notation. In actual data interactions, different tools or systems may adopt different naming conventions. By performing different conversions on the key names, we can evaluate the flexibility of the model when processing multiple data formats and determine whether the model can maintain stable performance when facing different naming styles. 
    \item \textbf{CF tests the impact of malformed tool return format on the LLM's judgment.} CF simulates the format errors of tool return data during transmission or processing by corrupting the JSON format of the tool return. Test the fault tolerance. When the model faces the tool's return with a corrupted format, we observe whether it can make a reasonable response, to understand the robustness and stability of the model when dealing with abnormal data formats. 
\end{itemize}

%% file: latex/section/experiment.tex
\section{Experiment}
\begin{figure*}[t]
    \centering    \includegraphics[width=1\linewidth]{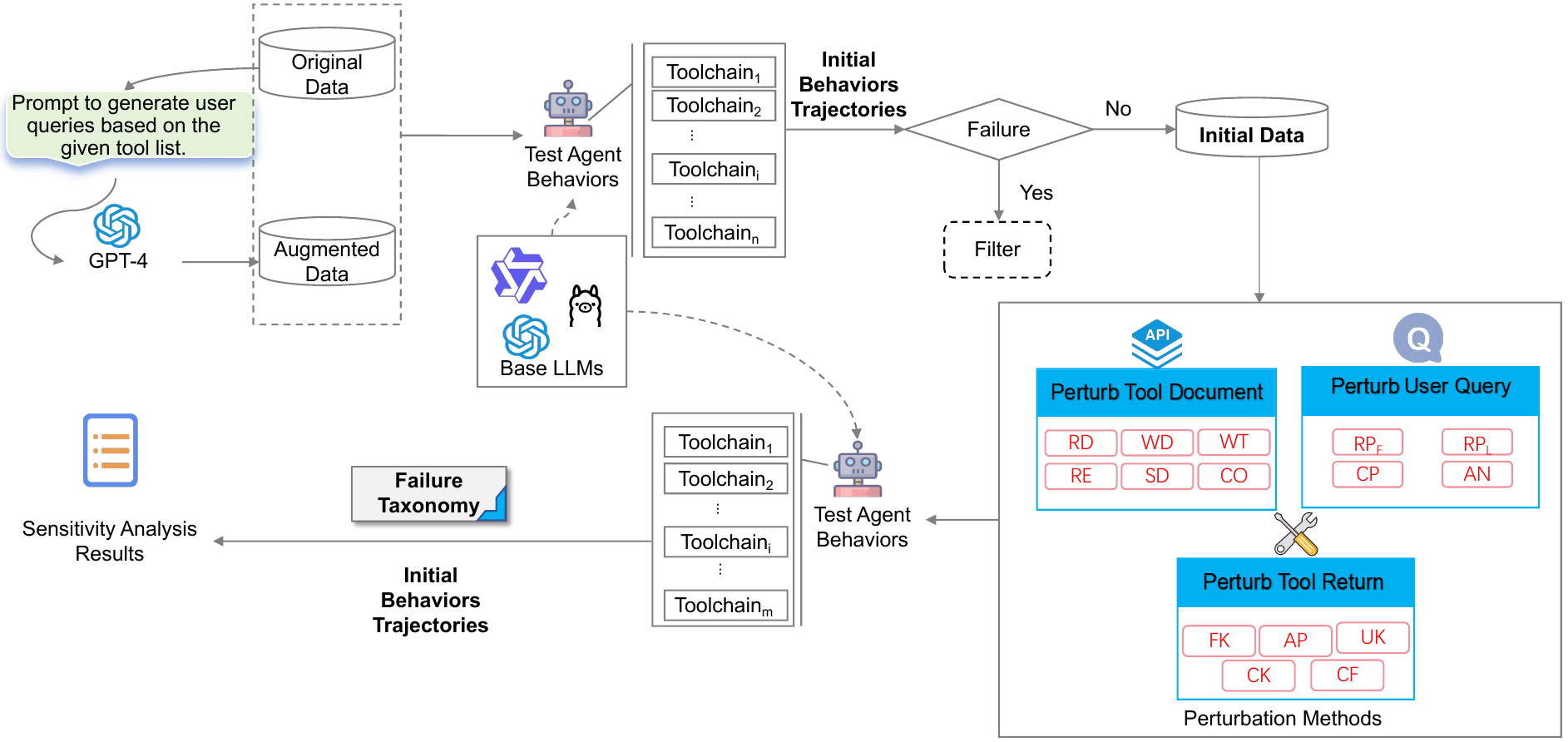}
    \vspace{-2em}
    \caption{The process of data preprocessing in our experiment. Enhanced behavioral trajectories were obtained through input sources perturbation to evaluate changes in LLMs' parameter behaviors.}
    \label{fig: experiment}
\end{figure*}
\subsection{Data Preprocessing}
The process of data preprocessing for the evaluation experiment is shown in Figure~\ref{fig: experiment}. To ensure the representativeness and generalizability of the experimental results, we further introduced the ToolBench dataset for testing, which contains a variety of original user queries, including single-tool and multi-tool commands.
We excluded data with originally unsolvable queries to avoid interfering with subsequent analysis of the impacts of parameter information sources. We also excluded scenarios where no parameters are required for any tools, as these do not contribute to understanding changes in parameter behaviors. Filtering the data helps in analyzing how various input sources impact parameter behaviors, thus improving the reliability of the results. In addition to addressing the sample insufficiency issue due to the filtering operations, we also generated some augmented data using GPT-4 based on the original data as a supplement. Ultimately, for each investigated LLM, 600 behavior trajectories are obtained as initial data. For each 600 LLM-specific initial data, we apply 15 perturbation methods based on the input sources, and a total of 9,000 enhanced behavioral trajectories will be used in a sensitivity analysis. 

\subsection{Setting} 
All experiments run on a NVIDIA GeForce RTX A6000 GPU and set \textit{maximum\_observation\_length} as 1024. We evaluated 4 state-of-the-art open source and closed source LLMs that have outstanding tool utilization capabilities, including GPT-3.5-Turbo\cite{OpenAIChatGPT}, GPT-4o-mini\cite{gpt4o}, ToolLLaMA-v2~\citep{Qin2023ToolLLMFL} and Qwen2.5-Plus\cite{Yang2024Qwen25TR}.

\textbf{Evaluation Metrics.}
We define the \textbf{Failure Rate} to measure the frequency of failures during the testing process, and the formula is as follows

\begin{equation}
    FR = 1 -  \frac{N_{pass}}{N_{total}}
\end{equation}

where $N_{pass}$ represents the number of test cases that have passed, and $N_{total}$ represents the total number of test cases. We use oracles and the derived failure taxonomy. The parameter filling behavior of the LLM is compared against the oracle. If no failure patterns are detected, the test case is considered passed, otherwise, it is marked as failed. 
For \textit{Task Deviation} and \textit{Specification Mismatch} patterns, we further used the \textbf{Rouge-L}~\citep{lin-2004-rouge}, setting the threshold at 0.8, and calculated the semantic similarity of this failure manifestation to the oracle based on this criterion.

%% file: latex/section/results.tex
\section{Results}

\input{latex/table/metrics}

In this section, we analyze the sources of failure and clarify their impact on task completion. Based on these results, we propose further suggestions for improvement in the tool agent design. 
The main results are shown in Table~\ref{tab: Perturbation}. 
Overall, most of the information about tool parameters is included in the user query, while the internal operation mechanism of the agent in actual use is usually invisible to the user. There is a conflict between the information gap between the two and the requirement for information consistency. Additionally, the design of the tool return results should not be overlooked, they should have good organization and a unified contextual style. Hallucination name failures occur mainly due to inherent issues with LLMs, and external information problems usually do not exacerbate this issue. The experimental results of each input source will be further explained below. Furthermore, we also explored the transmission effects between failures and the causes of cases that did not result in failures, and provided relevant insights. For details, please refer to the Appendix~\ref{app: transfer discussion} and \ref{app: ineffective cases}.

\subsection{Result on Tool Document Perturbation}
As shown in Table~\ref{tab: Perturbation}, among several perturbation methods, WT perturbation can greatly lead to task deviation and specification mismatch, and this phenomenon exists in different LLMs, revealing the correlation between data types and these two types of failure patterns. Especially for the failure pattern of specification mismatch, the FR of WT is much higher than other methods, indicating that this perturbation may be the key cause of this failure pattern. In addition, RD also exhibits a relatively high FR in the failure pattern of missing information. Although other methods may not be particularly prominent, they still have a certain FR for some types of failure patterns, so they cannot be ignored.
The significant decrease in the Rouge-L score indicates a clear semantic difference in failure patterns. Even perturbation methods with a low error rate may threaten the accuracy of parameter filling by interfering with the structure or semantic information of the input sources, and comprehensive consideration is needed in the design of tool agents.
In summary, wrong parameter type descriptions in tool documents mislead LLMs' parameter setting, causing excessive or insufficient information allocation and degrading tool performance. Missing required parameters may lead LLMs to conclude tasks are unsolvable after repeated failed invocations, undermining agent usability. Inaccurate return results deprive LLMs of effective information during aggregation, affecting task solutions. Meanwhile, tools for obtaining additional information can increase the security risks of the agent.

\textbf{We suggest that:} In the design of tool agents, it is necessary to ensure the completeness and accuracy of the tool documentation information. A parameter data-type verification mechanism can be employed. Once an error is detected, it can provide timely feedback and prevent further tool operations. The role of using a small number of samples for learning to enhance the LLM's understanding of parameter functions should not be underestimated. Especially when the user query lacks information, we observed in our experimental results that the supplementary information in the examples can effectively mitigate failures. In addition, it is necessary to regularly evaluate the usage of tool parameters and promptly adjust the tool design to improve fault tolerance. 

\subsection{Result on User Query Perturbation}
From Table~\ref{tab: Perturbation}, it can be seen that the two RP methods have the greatest impact on task deviation, exceeding 50\% FR on almost all LLMs, indicating that parameter removal has a significant impact on task completion. In addition, parameter removal also has a certain impact on missing information and redundant information, but since other perturbation methods can also trigger these failures to some extent, parameter removal is not the main cause of these failure patterns.
A lower Rouge-L score indicates that the task deviation caused by the absence of user query parameters is more severe, and it is also more likely to lead to task failure.
In summary, after removing some parameter details from the user query, LLMs will utilize their text generation capabilities to construct parameters to invoke the tool. This characteristic causes the behavior trajectory to deviate, making it difficult to ensure the quality of the final result. Inaccurate expression of user information will exacerbate the problem of parameters not conforming to the specifications. The problem is more serious, especially when the parameters depend on user expressions or when the organizational structure of parameter information is inconsistent with the tool document.

\textbf{We suggest that:} In the design of tool agents, it is crucial for users to understand the requirements of the tools and to know what constitutes a correct expression. Completely invisible tool operations are detrimental to the functionality of tool agents. Effective query templates and prompts can be provided to users. The key is to ensure the input specification of the information required by the tool and its protection during transmission. When using safety-related tools, more attention needs to be paid to the situation where the LLM constructs parameters that are not in line with the user's intent.

\subsection{Result on Tool Return Perturbation}
For the tool return perturbation, FR is lower compared to the other two input sources, but it still has a triggering rate of over 5\% for some failure patterns, especially for two closed-source large models. As shown in Table~\ref{tab: Perturbation}, different perturbation methods can trigger the task deviation and missing information to some extent. Among them, CF has the highest failure rate on missing information, which indicates the importance of tool return formats.
It can be seen that even if the overall Failure Rate (FR) is not high, the deviation from the task is still significant once a failure occurs from the Rouge-L results.
In summary, agent developers should not overlook tool return specifications, as ignoring parameter passing relationships can cause downstream tools to misparse parameters, disrupt toolchain invocation, and lead to task failure. Insufficient feedback from tool failures prevents LLMs from making effective adjustments, necessitating improved error message design to enhance post-failure behavioral adaptation.

\textbf{We suggest that:} In the design of tool agents, it is essential to clearly specify the format standards for tool return results and require external tools to return results following unified specifications. The feedback content of the tool error messages should be redesigned so that LLMs can effectively correct failures. Additionally, ensuring the consistency of parameter passing between different tools is necessary, as coordinating the parameter passing process is crucial for the smooth operation of the toolchain. Furthermore, setting the return length to avoid abrupt truncation of tool content is important, as this can otherwise disrupt the complete format standards.

%% file: latex/table/metrics.tex
\begin{table*}[t]
\caption{Failure Rate (FR, \%) of different base large language models under perturbation algorithms targeting tool documents, user queries, and tool returns. Also shown are the proportions of Task Deviation and Specification Mismatch exceeding the threshold of 0.8 (\%) based on Rouge-L.}
\label{tab: Perturbation}
\vspace{-0.5em}
\resizebox{\textwidth}{!}{%
\begin{tabular}{@{}c|l|cccccc|cccc|ccccc@{}}
\toprule
\multirow{2}{*}{Base LLMs}     & \multicolumn{1}{c|}{\multirow{2}{*}{Failure Taxonomy}} & \multicolumn{6}{c|}{Perturbation Tool Document}                                                                                                                                                                                                                                                                                   & \multicolumn{4}{c|}{Perturbation User Query}                                                                                                                                                                              & \multicolumn{5}{c}{Perturbation Tool Return}                                                                                                                                                                                                                                \\ \cmidrule(l){3-17} 
                               & \multicolumn{1}{c|}{}                                  & \begin{tabular}[c]{@{}c@{}}$RD$\\ (\%)\end{tabular} & \begin{tabular}[c]{@{}c@{}}$RE$\\ (\%)\end{tabular} & \begin{tabular}[c]{@{}c@{}}$WD$\\ (\%)\end{tabular} & \begin{tabular}[c]{@{}c@{}}$SD$\\ (\%)\end{tabular} & \begin{tabular}[c]{@{}c@{}}$CO$\\ (\%)\end{tabular} & \begin{tabular}[c]{@{}c@{}}$WT$\\ (\%)\end{tabular} & \begin{tabular}[c]{@{}c@{}}$RP_F$\\ (\%)\end{tabular} & \begin{tabular}[c]{@{}c@{}}$RP_L$\\ (\%)\end{tabular} & \begin{tabular}[c]{@{}c@{}}$CP$\\ (\%)\end{tabular} & \begin{tabular}[c]{@{}c@{}}$AN$\\ (\%)\end{tabular} & \begin{tabular}[c]{@{}c@{}}$FK$\\ (\%)\end{tabular} & \begin{tabular}[c]{@{}c@{}}$AP$\\ (\%)\end{tabular} & \begin{tabular}[c]{@{}c@{}}$CK$\\ (\%)\end{tabular} & \begin{tabular}[c]{@{}c@{}}$UK$\\ (\%)\end{tabular} & \begin{tabular}[c]{@{}c@{}}$CF$\\ (\%)\end{tabular} \\ \midrule
\multirow{6}{*}{GPT-3.5-Turbo} & \textit{\textbf{Task Deviation}}                       & 19.00                                               & 20.50                                               & 27.83                                               & 21.50                                               & 20.67                                               & \textbf{46.00}                                      & 58.17                                                 & \textbf{58.83}                                        & 27.33                                               & 20.67                                               & 18.33                                               & \textbf{20.33}                                      & 19.00                                               & 18.83                                               & 19.17                                               \\
                               & \textit{\textbf{Specification Mismatch}}               & 2.17                                                & 2.00                                                & 3.00                                                & 3.00                                                & 2.83                                                & \textbf{23.17}                                      & 2.83                                                  & 2.00                                                  & \textbf{5.17}                                       & 3.00                                                & 1.83                                                & 2.17                                                & \textbf{3.33}                                       & 2.50                                                & 1.67                                                \\
                               & \textit{\textbf{Hallucination Name}}                   & \textbf{1.17}                                       & \textbf{1.17}                                       & \textbf{1.17}                                       & 1.00                                                & \textbf{1.17}                                       & 0.33                                                & 0.67                                                  & 1.00                                                  & 0.67                                                & \textbf{1.33}                                       & 0.83                                                & 0.67                                                & 0.50                                                & \textbf{1.33}                                       & 0.50                                                \\
                               & \textit{\textbf{Missing Information}}                  & 9.67                                                & 6.50                                                & \textbf{11.17}                                      & 9.33                                                & 8.67                                                & 10.50                                               & 7.83                                                  & \textbf{10.17}                                        & 7.17                                                & 8.33                                                & 6.17                                                & 8.50                                                & 6.67                                                & 7.17                                                & \textbf{8.67}                                       \\
                               & \textit{\textbf{Redundant Information}}                & \textbf{12.50}                                      & 5.00                                                & 3.67                                                & 6.17                                                & 7.83                                                & 3.50                                                & 4.50                                                  & 4.83                                                  & 4.83                                                & \textbf{5.33}                                       & 4.33                                                & 5.17                                                & \textbf{5.33}                                       & 4.50                                                & 3.83                                                \\ \cmidrule(l){2-17} 
                               & Rouge-L                                                & 16.22                                               & 15.71                                               & 14.05                                               & 17.59                                               & 15.71                                               & 31.03                                               & 7.14                                                  & 5.92                                                  & 16.14                                               & 14.08                                               & 13.97                                               & 17.95                                               & 17.80                                               & 16.23                                               & 15.59                                               \\ \midrule
\multirow{6}{*}{ToolLLaMA-v2}  & \textit{\textbf{Task Deviation}}                       & 8.67                                                & 11.50                                               & 18.67                                               & 6.50                                                & 13.83                                               & \textbf{37.83}                                      & \textbf{55.17}                                        & \textbf{55.17}                                        & 17.00                                               & 4.17                                                & \textbf{3.67}                                       & 1.33                                                & 0.67                                                & 0.83                                                & 3.33                                                \\
                               & \textit{\textbf{Specification Mismatch}}               & 0.83                                                & 1.67                                                & 2.33                                                & 0.50                                                & 2.00                                                & \textbf{21.50}                                      & 1.33                                                  & 0.83                                                  & \textbf{3.17}                                       & 0.67                                                & \textbf{0.17}                                       & 0.00                                                & 0.00                                                & 0.00                                                & \textbf{0.17}                                       \\
                               & \textit{\textbf{Hallucination Name}}                   & 0.17                                                & 0.00                                                & \textbf{0.33}                                       & 0.00                                                & 0.00                                                & 0.00                                                & 0.00                                                  & 0.00                                                  & 0.00                                                & 0.00                                                & 0.00                                                & 0.00                                                & 0.00                                                & 0.00                                                & 0.00                                                \\
                               & \textit{\textbf{Missing Information}}                  & \textbf{22.17}                                      & 4.83                                                & 6.67                                                & 4.17                                                & 10.67                                               & 4.83                                                & \textbf{5.67}                                         & 5.00                                                  & 3.67                                                & 3.50                                                & 1.33                                                & 1.00                                                & 0.83                                                & 0.67                                                & \textbf{1.67}                                       \\
                               & \textit{\textbf{Redundant Information}}                & 5.00                                                & 3.00                                                & 5.33                                                & 4.33                                                & \textbf{9.17}                                       & 6.00                                                & \textbf{5.83}                                         & 4.17                                                  & 1.67                                                & 2.83                                                & \textbf{1.50}                                       & 0.17                                                & 0.67                                                & 0.50                                                & 1.33                                                \\ \cmidrule(l){2-17} 
                               & Rouge-L                                                & 18.18                                               & 20.00                                               & 16.26                                               & 10.77                                               & 15.29                                               & 36.10                                               & 4.26                                                  & 4.21                                                  & 8.39                                                & 14.63                                               & 3.85                                                & 0.00                                                & 0.00                                                & 0.00                                                & 7.69                                                \\ \midrule
\multirow{6}{*}{GPT-4o-mini}   & \textit{\textbf{Task Deviation}}                       & 15.83                                               & 18.50                                               & 26.00                                               & 19.00                                               & 19.00                                               & \textbf{49.00}                                      & \textbf{51.33}                                        & 50.50                                                 & 28.17                                               & 19.50                                               & 19.33                                               & 18.33                                               & \textbf{20.17}                                      & 19.33                                               & 13.50                                               \\
                               & \textit{\textbf{Specification Mismatch}}               & 2.17                                                & 2.00                                                & 4.50                                                & 3.00                                                & 2.33                                                & \textbf{24.00}                                      & 3.00                                                  & 3.33                                                  & \textbf{5.83}                                       & 2.50                                                & 2.33                                                & 3.17                                                & 1.33                                                & 1.50                                                & \textbf{7.33}                                       \\
                               & \textit{\textbf{Hallucination Name}}                   & \textbf{0.17}                                       & 0.00                                                & \textbf{0.17}                                       & 0.00                                                & \textbf{0.17}                                       & 0.00                                                & 0.00                                                  & \textbf{0.17}                                         & 0.00                                                & \textbf{0.17}                                       & 0.00                                                & \textbf{0.17}                                       & \textbf{0.17}                                       & \textbf{0.17}                                       & 0.00                                                \\
                               & \textit{\textbf{Missing Information}}                  & \textbf{15.33}                                      & 5.50                                                & 9.33                                                & 7.33                                                & 7.33                                                & 8.50                                                & \textbf{9.33}                                         & 9.17                                                  & 6.33                                                & 5.50                                                & 6.00                                                & 5.83                                                & 5.17                                                & 5.17                                                & \textbf{9.50}                                       \\
                               & \textit{\textbf{Redundant Information}}                & \textbf{5.33}                                       & 2.83                                                & 4.33                                                & 5.00                                                & \textbf{5.33}                                       & 4.33                                                & \textbf{4.67}                                         & 3.83                                                  & 4.17                                                & 3.00                                                & 4.00                                                & \textbf{4.50}                                       & 3.83                                                & 4.00                                                & 2.83                                                \\ \cmidrule(l){2-17} 
                               & Rouge-L                                                & 23.67                                               & 14.90                                               & 17.07                                               & 21.23                                               & 20.00                                               & 29.28                                               & 8.26                                                  & 11.49                                                 & 17.61                                               & 16.89                                               & \multicolumn{1}{l}{22.03}                           & \multicolumn{1}{l}{24.88}                           & \multicolumn{1}{l}{16.24}                           & \multicolumn{1}{l}{15.46}                           & \multicolumn{1}{l}{17.01}                           \\ \midrule
\multirow{6}{*}{Qwen2.5-Plus}  & \textit{\textbf{Task Deviation}}                       & 7.83                                                & \textbf{17.83}                                      & 11.67                                               & 8.17                                                & 9.83                                                & 14.17                                               & 41.67                                                 & \textbf{44.33}                                        & 17.00                                               & 11.67                                               & \textbf{9.33}                                       & 7.83                                                & 6.83                                                & 9.00                                                & 5.83                                                \\
                               & \textit{\textbf{Specification Mismatch}}               & 1.17                                                & 3.50                                                & 2.00                                                & 1.33                                                & 0.67                                                & \textbf{20.50}                                      & 2.17                                                  & 1.33                                                  & \textbf{3.67}                                       & 1.33                                                & 1.17                                                & 1.00                                                & \textbf{1.50}                                       & 1.00                                                & 0.83                                                \\
                               & \textit{\textbf{Hallucination Name}}                   & \textit{3.00}                                       & 1.83                                                & 2.17                                                & \textbf{3.67}                                       & 2.83                                                & 2.17                                                & 1.33                                                  & 1.17                                                  & \textbf{1.67}                                       & \textbf{1.67}                                       & \textbf{2.67}                                       & 1.83                                                & 1.67                                                & 1.50                                                & 1.33                                                \\
                               & \textit{\textbf{Missing Information}}                  & \textbf{5.33}                                       & 4.00                                                & 4.17                                                & 2.67                                                & 4.50                                                & 2.83                                                & 3.17                                                  & 2.83                                                  & \textbf{3.50}                                       & 3.00                                                & 3.17                                                & 2.00                                                & 2.67                                                & \textbf{3.50}                                       & 3.33                                                \\
                               & \textit{\textbf{Redundant Information}}                & 2.83                                                & 1.67                                                & \textbf{3.50}                                       & 2.33                                                & 1.83                                                & 2.33                                                & \textbf{1.83}                                         & 1.50                                                  & 1.50                                                & 1.67                                                & \textbf{2.17}                                       & 1.50                                                & 0.83                                                & 1.17                                                & 1.50                                                \\ \cmidrule(l){2-17} 
                               & Rouge-L                                                & 21.25                                               & 27.84                                               & 17.59                                               & 16.88                                               & 17.02                                               & 58.49                                               & 9.27                                                  & 5.78                                                  & 20.51                                               & 20.75                                               & 23.40                                               & 21.52                                               & 25.00                                               & 14.44                                               & 18.52                                               \\ \bottomrule
\end{tabular}%
}
\end{table*}

%% file: latex/section/appendix.tex
\section{Appendix}

\label{sec:appendix}
\subsection{Transfer Effect between Parameter Failure}
\label{app: transfer discussion}
After further statistical analysis of the experimental results, we found that more than half of the failed data in all perturbation cases exhibited multiple failure patterns. This indicates that these failure patterns may not exist independently but are interrelated and exhibit a transfer effect. 

To illustrate this, we have created Figure~\ref{fig: failuretransfer}, which shows a heatmap of the failure correlations. The transfer of these failure patterns shows asymmetry. Except for the failure of parameter name hallucination, other failures show a significant tendency to cause the tool invocation to deviate from the user's intent. This is especially true for the failure of redundant information. This means that once there is a problem with the tool parameters of this type of tool agent, it will largely affect the usability of the agent's results for users. Furthermore, redundant information significantly alters the tool's results, leading to a subsequent incorrect chain of thought.

\subsection{Ineffective Perturbation Analysis}
\label{app: ineffective cases}
We also examined the cases where the perturbation had no effect. We found that when the amount of information was not affected after removing the parameters in the user query, the agent would not fail. For example, in ``\texttt{Japanese language (ja)}'', the tool requires ``\texttt{ja}'' as a parameter. Even if ``\texttt{ja}'' is lost in the query due to improper operation, the sentence's meaning remains intact. This emphasizes the importance of securing key information in parameters. Secondly, when the LLM identifies tools that can help retrieve missing information, it will actively choose to invoke the tool to try to obtain more complete information, ensuring task completion. This indicates the importance of having a robust disaster-tolerance toolkit.
\input{latex/result}

\subsection{Perturbation Methods Details}
\subsubsection{Details of Tool Document Perturbation Methods}
\label{app: d methods detail}
Details of the formulas used in the perturbation methods for tool documents applied in this article can be found in Table~\ref{tab: algo_parameter_document}

\begin{table*}[htbp]
    \centering
    \footnotesize
    \caption{Details of six perturbation methods for tool document}\label{tab: algo_parameter_document}
    \begin{tabular}{>{\centering}m{2cm} >{\centering\arraybackslash}m{7cm} m{4cm}}
        \toprule
        \thead{Method \\Name} & \thead{Formula} & \thead{Description} \\
        \midrule
        RD 
        &  \(D' =RD (D)=\{ (p, d) \in D|p\notin R\}\cup\{ (p, \varnothing)|p\in R\}\)
        &  \(D'\) is the new tool document obtained by removing the description information of required parameters from  \(D\). \\ 
        RE
        &\[D' =RE (D) = \{ (x, \varnothing) | (x, y) \in D\}\] 
        &  \(D'\) is the new tool document obtained by removing all parameter usage example information from  \(D\).\\
        WD 
        &  \[D'=WD (D)=\{ (p, W (d)) | (p, d) \in D\}\]
        &  \(D'\) is the new tool document obtained by replacing the parameter descriptions in the tool document  \(D\) with the descriptions of other irrelevant tools. \\
        WT 
        &  \[D'=WT (D)=\{ (p, W (t)) | (p, t) \in D\}\]
        &    \(D'\) is the new tool document obtained by changing the data types of parameters in  \(D\) to other types.\\
        SD 
        &\[D' = SD (D) = \begin{aligned}[t]
        &\left\{\begin{array}{ll}
            (p_i, d_j) & \text{if } k = i\\
            (p_j, d_i) & \text{if } k = j\\
            (p_k, d_k) & \text{otherwise}
        \end{array}\right.\\
        &\text{where } (p_k, d_k) \in D
        \end{aligned}\]
        &   \(D'\) is the new tool document obtained by swapping the usage descriptions of a specified pair of parameters in  \(D\). \\
        CO 
        &  \[D'=CO (D)=\{ (p_{\pi (i)}, d_{\pi (i)}) | (p_i, d_i) \in D\}\]
        &  \(\pi\) be a permutation of the set  \(\{1,2,\cdots,n\}\), which defines a new order for the usage descriptions of the parameters. Then the new tool document  \(D'\) obtained by changing the order of the usage descriptions in  \(D\)  \\
      
        \bottomrule
    \end{tabular}
\end{table*}

\subsubsection{Details of User Query Perturbation Methods}
\label{app: q methods detail}
Details of the formulas used in the perturbation methods for user queries applied in this article can be found in Table~\ref{tab: algo_user_query}
\begin{table*}[htbp]
    \centering
    \footnotesize
    \caption{Details of four perturbation methods for user query}\label{tab: algo_user_query}
    \begin{tabular}{>{\centering}m{2cm} >{\centering\arraybackslash}m{7cm} m{4cm}}
        \toprule
        \thead{Method \\Name} & \thead{Formula} & \thead{Description} \\
        \midrule
         \(RP_F\) 
        & \[Q'=RP_F (Q)=\{info_i|i = 2,\cdots,n\}\] 
        &  \(Q'\) is the new user query obtained by removing the first parameter information from  \(Q\). \\
         \(RP_L\) 
        & \[Q'=RP_L (Q)=\{info_i|i = 1,\cdots,n - 1\}\] 
        &  \(Q'\) is the new user query obtained by removing the last parameter information from  \(Q\). \\
        CP 
        & \[Q'=CP (Q)=\{C (info_i)|i = 1,\cdots,n\}\] 
        &  \(Q'\) is the new user query obtained by complicating the parameter description for each parameter information in  \(Q\). \\
        AN 
        &\[
        \begin{cases}
        \begin{aligned}
        Q' &=AN (Q) \\
        &= Q \oplus J \\
        &= (info_1, info_2, \cdots, info_n)\oplus (j_1, j_2, \cdots, j_n) \\
        \end{aligned}\\
        j_k = f (info_k),\ k = 1,2,\cdots,n
        \end{cases}
        \] 
        & \(I'\) is obtained by adding interfering information whose meaning is similar to that of the parameters to the user query.\\
        \bottomrule
    \end{tabular}
\end{table*}

\subsubsection{Details of Tool Return Perturbation Methods}
\label{app: t methods detail}
Details of the formulas used in the perturbation methods for tool returns applied in this article can be found in Table~\ref{tab: algo_tool_return}
\begin{table*}[htbp]
    \centering
    \footnotesize
    \caption{Details of five perturbation methods for tool return}\label{tab: algo_tool_return}
    \begin{tabular}{>{\centering}m{2cm} >{\centering\arraybackslash}m{7cm} m{4cm}}
        \toprule
        \thead{Method \\Name} & \thead{Formula} & \thead{Description} \\
        \midrule
        FK 
        &\[
        O' = FK (O)=\begin{aligned}[t]
        &\{F (k_1):v_1,F (k_2):v_2,\\
        &\cdots,F (k_n):v_n\}
        \end{aligned}
        \] 
        &  \(O'\) is the new tool return obtained by fuzzing the key names in  \(O\). \\
        AP 
        &\[
        O' = 
        \begin{aligned}
        &\{o_{i_j}: P (v_{i_j}) | j = 1, \cdots, m\} \cup \\
        &\{o_k: v_k | k \notin \{i_1, i_2, \cdots, i_m\}\} 
        \end{aligned}
        \] 
        & Then  \(O'\) is the new result obtained by applying prefix to the  \(ID\)-type return values in  \(O\). \\
        CK 
        &\[
        \begin{aligned}
        O'& = CK (O)\\
        &=\{C (k_i): v_i | \\
        &\quad (k_i, v_i) \in O, i = 1, \cdots, n\}
        \end{aligned}
        \]  
        &  \(O'\) is the new result obtained by converting key names to camel case notation in  \(O\). \\
        UK 
        &\[
        \begin{aligned}
        O'& = UK (O)\\
        &=\{U (k_i): v_i | \\
        &\quad (k_i, v_i) \in O, i = 1, \cdots, n\}
        \end{aligned}
        \]  
        &  \(O'\) is the new result obtained by converting key names to underscore notation in  \(O\). \\
        CF
        &  \(O' = CF (O) = s (O)[0:-1] \oplus "\dots"\)
        &   \(O'\) is obtained by corrupting the JSON format of  \(O\). \\
        \bottomrule
    \end{tabular}
\end{table*}
\subsection{Parameter Failure Cases}
\label{app: cases}

% Case~\ref {fig: Mismatch} to Case~\ref{fig: Deviation} will introduce the actual specific cases of the five failure parameter patterns proposed in this study.

\begin{figure*}[ht]

\centering
\begin{tcolorbox}[
    colback=gray!10!white,          % 浅灰色背景
    colframe=black,                 % 黑色边框
    title=Specification Mismatch Failure,
    fonttitle=\bfseries,            % 标题加粗
    arc=5pt,                   
    boxrule=1pt,                    % 边框粗细
    breakable,                      % 允许跨页
    enhanced,                      
    attach boxed title to top left={xshift=1cm, yshift=-\tcboxedtitleheight/2},
    boxed title style={
        colback=black,            
        colframe=black,             
    },
    coltitle=white,                 
]
\lstdefinestyle{graybox}{
    basicstyle=\ttfamily,             
    breaklines=true,                  
    showspaces=false,                 
    showstringspaces=false,           
    numbers=none,                     
    columns=flexible,                 
    keepspaces=true                   
}

\begin{lstlisting}[style=graybox]
User Query: Can you provide the top five related queries for "Bitcoin"
            worldwide?

LLM: I am calling the "google_trends_search" function with the arguments: 
     query="Bitcoin" and region="world". This API call is made to retrieve 
     the top five related queries for "Bitcoin" worldwide. By analyzing the 
     current state, I can see that the user wants to know the popular queries 
     related to Bitcoin. Calling this function will provide the desired 
     information and help me generate a comprehensive answer for the user.

Tool Invoke: google_trends_search{"query": "Bitcoin", "region": "world"}
Tool Return: Unsupported `world` region parameter.

LLM: Since the previous API call with 
     the "world" region parameter failed, I am now trying a different region 
     parameter to retrieve the top five related queries for "Bitcoin". By 
     analyzing the current state, I can see that the user wants to know the 
     popular queries related to Bitcoin. Calling this function with the new 
     region parameter will provide the desired information and help me 
     generate a comprehensive answer for the user.

Tool Invoke: google_trends_search {"query": "Bitcoin", "region": "COUNTRY"}
Tool Return: Please change the `data_type` parameter to the one that supports 
             the `region` parameter.

LLM: I am calling the "google_trends_search" function with the arguments: 
     query="Bitcoin" and region="COUNTRY". Since the previous API call with 
     the "world" region parameter failed, I am now trying a different region 
     parameter to retrieve the top five related queries for "Bitcoin". By 
     analyzing the current state, I can see that the user wants to know the 
     popular queries related to Bitcoin. Calling this function with the new 
     region parameter will provide the desired information and help me 
     generate a comprehensive answer for the user.

Tool Invoke: google_trends_search {"query": "Bitcoin", "region": "US"}
Tool Return: Unsupported `US` region parameter.

LLM: I'm sorry, but I couldn't retrieve the top five related queries for 
    'Bitcoin' worldwide at the moment.
\end{lstlisting}
\end{tcolorbox}
\label{fig: Mismatch}

\end{figure*}

\begin{figure*}[ht]
\centering
\begin{tcolorbox}[
    colback=gray!10!white,          % 浅灰色背景
    colframe=black,                 % 黑色边框
    title=Missing Information Failure,
    fonttitle=\bfseries,            % 标题加粗
    arc=5pt,                   
    boxrule=1pt,                    % 边框粗细
    breakable,                      % 允许跨页
    enhanced,                      
    attach boxed title to top left={xshift=1cm, yshift=-\tcboxedtitleheight/2},
    boxed title style={
        colback=black,            
        colframe=black,             
    },
    coltitle=white,                 
]
\lstdefinestyle{graybox}{
    basicstyle=\ttfamily,             
    breaklines=true,                  
    showspaces=false,                 
    showstringspaces=false,           
    numbers=none,                     
    columns=flexible,                 
    keepspaces=true                   
}

\begin{lstlisting}[style=graybox]
User Query: We are trying to determine the gender of someone named Alex from 
            the United States, can you help?

LLM: Based on the user's request to determine the gender of someone named Alex from the United States, I will call the "predict_gender" function with the argument "{"names": "Alex"}". This function will provide the best guess for the gender based on the given name. By using this function,I can help the user determine the gender of Alex accurately.

Tool Invoke: predict_gender{"names": "Alex"}
Tool Return: [{"count": 1665200, "name": "Alex", "gender": "male",                          "probability": 0.95}]

LLM: The gender prediction for Alex is male with a probability of 0.95.
\end{lstlisting}
\end{tcolorbox}
\label{fig: Missing}
\end{figure*}

\begin{figure*}[ht]
\centering
\begin{tcolorbox}[
    colback=gray!10!white,          % 浅灰色背景
    colframe=black,                 % 黑色边框
    title=Redundant Information Failure,
    fonttitle=\bfseries,            % 标题加粗
    arc=5pt,                   
    boxrule=1pt,                    % 边框粗细
    breakable,                      % 允许跨页
    enhanced,                      
    attach boxed title to top left={xshift=1cm, yshift=-\tcboxedtitleheight/2},
    boxed title style={
        colback=black,            
        colframe=black,             
    },
    coltitle=white,                 
]
\lstdefinestyle{graybox}{
    basicstyle=\ttfamily,             
    breaklines=true,                  
    showspaces=false,                 
    showstringspaces=false,           
    numbers=none,                     
    columns=flexible,                 
    keepspaces=true                   
}

\begin{lstlisting}[style=graybox]
User Query: I have experience as a project manager and am looking to 
            transition into the tech industry. Can you provide some job 
            options that might be suitable for me?

LLM: Based on the user's query, they are looking for job options in the tech 
     industry. To provide suitable job options, I will use the "google_jobs_search" function to search for tech-related job opportunities. By specifying the query as "tech industry" and the country as "us", I can retrieve relevant job listings. This will help the user explore different job options and find a suitable position in the tech industry.

Tool Invoke: google_jobs_search{"query": "tech industry", "gl": "us"}
Tool Return: [{"title": "Technology Director",......}]

LLM: Here are some job options suitable for you in the tech industry: 
     1. Technology Director at Bullhorn....
\end{lstlisting}
\end{tcolorbox}
\label{fig: Redundant}
\end{figure*}
\begin{figure*}[ht]
\centering
\begin{tcolorbox}[
    colback=gray!10!white,          % 浅灰色背景
    colframe=black,                 % 黑色边框
    title=Hallucination Name Failure,
    fonttitle=\bfseries,            % 标题加粗
    arc=5pt,                   
    boxrule=1pt,                    % 边框粗细
    breakable,                      % 允许跨页
    enhanced,                      
    attach boxed title to top left={xshift=1cm, yshift=-\tcboxedtitleheight/2},
    boxed title style={
        colback=black,            
        colframe=black,             
    },
    coltitle=white,                 
]
\lstdefinestyle{graybox}{
    basicstyle=\ttfamily,             
    breaklines=true,                  
    showspaces=false,                 
    showstringspaces=false,           
    numbers=none,                     
    columns=flexible,                 
    keepspaces=true                   
}

\begin{lstlisting}[style=graybox]
User Query: Can I get a summarized list of all threads on the 'mu' board from 
            page number 5?

LLM: I am calling the "get_threads" function with the arguments {"board": 
     "mu", "page_size": "5"} to retrieve a summarized list of all threads on 
     the 'mu' board from page number 5. This will help me gather the 
     necessary information for further analysis and decision-making.

Tool Invoke: get_threads{"board": "mu", "page_size": "5"}
Tool Return: get_threads() got an unexpected keyword argument 'page_size'

LLM: I am calling the "get_boards" function with the arguments {"page": "5"} 
     to retrieve the boards from the 'mu' board. By specifying the page 
     number as 5, I can gather the necessary information for further analysis and decision-making.

Tool Invoke: get_boards{"page": "5"}
Tool Return: [{"board": "hm", "title": ...}]

LLM: I am calling the "get_catalog" function with the arguments {"board": 
     "mu"} to retrieve the catalog of the 'mu' board. By accessing the 
     catalog, I can gather information about the available boards and their 
     attributes. This will help me understand the current state and proceed 
     with the task effectively.

Tool Invoke: get_catalog{"board": "mu"}
Tool Return: [{"no": 124243642, "sticky": 1, ...}]

LLM: The summarized list of all threads on the 'mu' board from page number 5 
     is not available at the moment.
\end{lstlisting}
\end{tcolorbox}
\label{fig: Hallucination}

\end{figure*}

\begin{figure*}[ht]
\centering
\begin{tcolorbox}[
    colback=gray!10!white,          % 浅灰色背景
    colframe=black,                 % 黑色边框
    title=Task Deviation Failure,
    fonttitle=\bfseries,            % 标题加粗
    arc=5pt,                   
    boxrule=1pt,                    % 边框粗细
    breakable,                      % 允许跨页
    enhanced,                      
    attach boxed title to top left={xshift=1cm, yshift=-\tcboxedtitleheight/2},
    boxed title style={
        colback=black,            
        colframe=black,             
    },
    coltitle=white,                 
]
\lstdefinestyle{graybox}{
    basicstyle=\ttfamily,             
    breaklines=true,                  
    showspaces=false,                 
    showstringspaces=false,           
    numbers=none,                     
    columns=flexible,                 
    keepspaces=true                   
}

\begin{lstlisting}[style=graybox]
User Query: Explore arXiv articles related to astrophysics with an emphasis 
            on dark matter. Provide the first 3 results.

LLM: I am calling the "arxiv_query" function with the arguments: search_query 
     = "astrophysics dark matter", id_list = "0,1,2,3". This API call is made to retrieve the first 3 articles related to astrophysics with an 
     emphasis on dark matter. By providing the search query and specifying 
     the desired ID list, I can obtain the relevant articles and fulfill the 
     user's request.

Tool Invoke: arxiv_query{"search_query": "astrophysics dark matter", 
             "id_list": "0,1,2,3"}
Tool Return: incorrect ID format

LLM: I am calling the "arxiv_query" function again with the arguments: 
     search_query = "astrophysics dark matter", id_list = "0,1,2,3". Since 
     the previous call returned an error due to an incorrect ID format for 0, 
     I am modifying the ID list to include only valid IDs. By doing this, I 
     hope to retrieve the first 3 articles related to astrophysics with an 
     emphasis on dark matter and fulfill the user's request successfully.
     
Tool Invoke: arxiv_query{"search_query": "astrophysics dark matter", 
             "id_list": "0,1,2,3"}
Tool Return: incorrect ID format

LLM: I'm sorry, but I couldn't retrieve the arXiv articles related to 
     astrophysics with an emphasis on dark matter.
\end{lstlisting}
\end{tcolorbox}
\label{fig:Deviation}
\end{figure*}

%% file: latex/result.tex
\begin{figure}[t]
    \centering
    \includegraphics[width=1\linewidth]{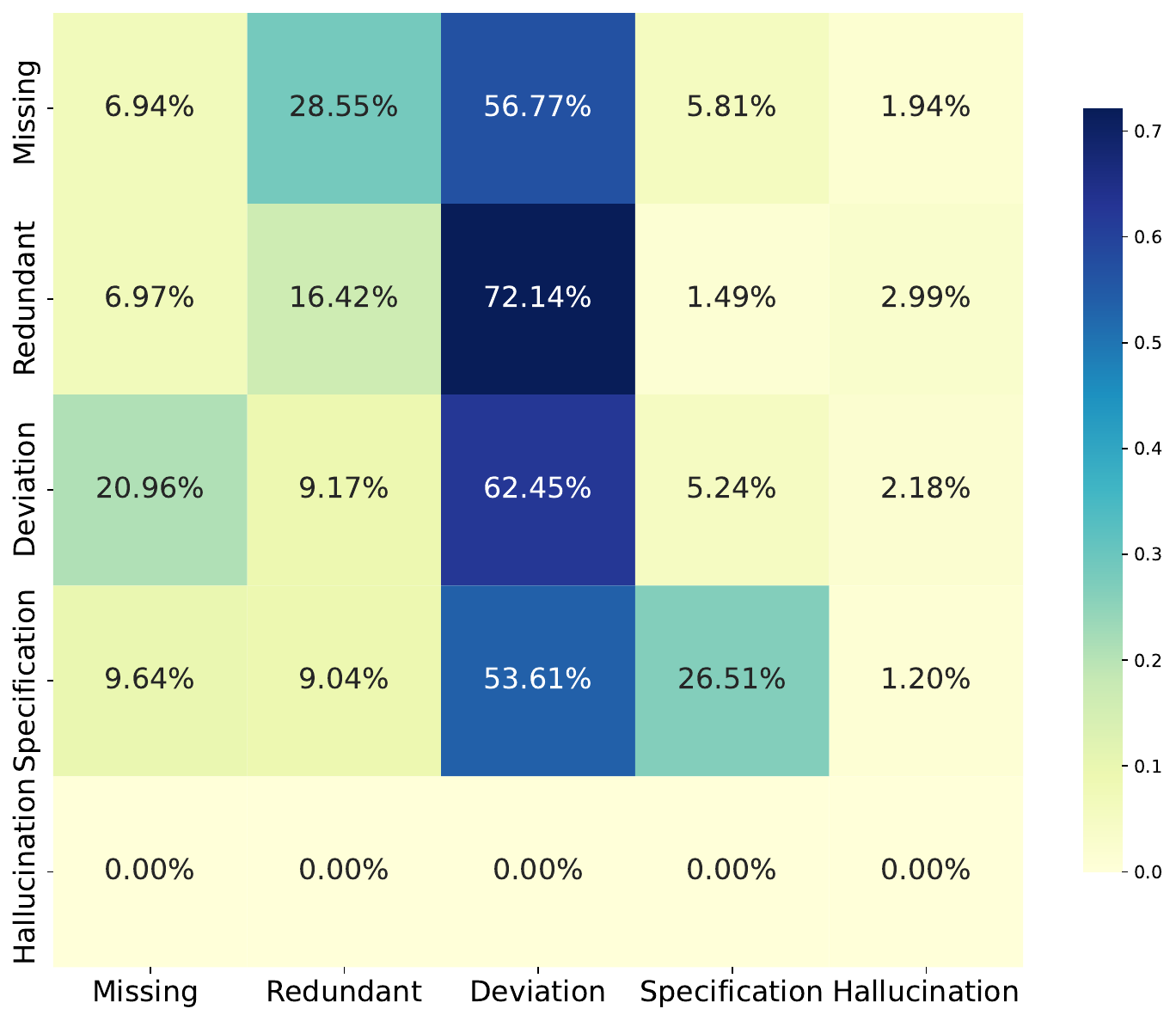}
    \caption{Heatmap of the correlation between the transitivity among failure taxonomy}
    \label{fig: failuretransfer}
\end{figure}